\documentclass[twocolumn,showpacs,amsmath,amssymb,pra,superscriptaddress,floatfix]{revtex4}
\usepackage{graphicx}
\usepackage{bm}
\usepackage{amsmath,amsfonts}

\begin{document}

\title{Fermionic collective excitations in a lattice gas of Rydberg atoms}

\author{B. Olmos}
\affiliation{Instituto 'Carlos I' de F\'{\i}sica Te\'orica y Computacional
and Departamento de F\'{\i}sica At\'omica, Molecular y Nuclear,
Universidad de Granada, E-18071 Granada, Spain}
\affiliation{Midlands Ultracold Atom Research Centre - MUARC, The University of Nottingham, School of Physics and Astronomy, Nottingham, United Kingdom}
\author{R. Gonz\'{a}lez-F\'{e}rez}
\affiliation{Instituto 'Carlos I' de F\'{\i}sica Te\'orica y Computacional
and Departamento de F\'{\i}sica At\'omica, Molecular y Nuclear,
Universidad de Granada, E-18071 Granada, Spain}
\author{I. Lesanovsky}
\email{igor.lesanovsky@nottingham.ac.uk}
\affiliation{Midlands Ultracold Atom Research Centre - MUARC, The University of Nottingham, School of Physics and Astronomy, Nottingham, United Kingdom}

\date{\today}

\begin{abstract}
We investigate the many-body quantum states of a laser-driven gas of Rydberg atoms confined to a large spacing ring lattice. If the laser driving is much stronger than the van-der-Waals interaction among the Rydberg sates, these many-body states are collective fermionic excitations. The first excited state is a spin-wave that extends over the entire lattice. We demonstrate that our system permits to study fermions in the presence of disorder although no external atomic motion takes place. We analyze how this disorder influences the excitation properties of the fermionic states. Our work shows a route towards the creation of complex many-particle states with atoms in lattices.
\end{abstract}
\pacs{32.80.Ee, 42.50.Dv, 67.85.-d, 71.45.-d}
\maketitle
Experiments with ultracold ground state atoms have opened a doorway to the study of many-particle systems and revealed deep insights into many-body physics and the dynamics of phase transitions \cite{Bloch08}. Entangled states of atomic ensembles serve as a resource for quantum information processing, precision measurement and the generation and storage of light \cite{Haroche06}. Recently, there is a growing interest in highly excited atomic (Rydberg) states as the strong state-dependent interaction between these atoms allows the implementation of fast quantum information protocols \cite{Jaksch00,Mueller09,Saffman09} and the creation of strongly interacting (spin) systems which show collective excitation behavior \cite{Heidemann07,MuramatsuPfau08}. The long-ranged character of the interaction allows to entangle atoms with a spatial separation of the order of several micrometers, as has recently been demonstrated experimentally \cite{Urban08,Gaetan08}.

\begin{figure}
\includegraphics[width=7cm]{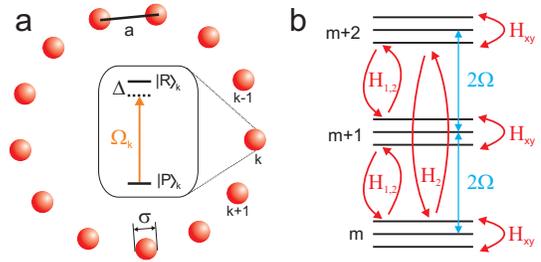}
\caption{\textbf{a}: Ring lattice with spacing $a$ being much larger than the extension $\sigma$ of the Wannier functions (deep lattice). The internal atomic degrees of freedom at each site are described by the (collective) states $\left|P\right>_k$ and $\left|R\right>_k$. \textbf{b}: Level structure in the regime in which $\Omega_k=\Omega\gg\beta$ and $|\Delta| \ll \Omega$. The spectrum splits into manifolds which can be labeled by the quantum number $m$ of the operator $\sum_k \sigma^{(k)}_z$. For sufficiently large $\Omega$, the coupling between manifolds established by $H_1$ and $H_2$ can be neglected. The (constrained) dynamics inside the $m$-subspaces is then determined by $H_\mathrm{xy}$. }\label{fig:spectrum}
\end{figure}

Motivated by this development we envisage atoms trapped in the sites of a deep large spacing \cite{Kinoshita06,Whitlock09} ring lattice with lattice constant $a\sim\mu m$ (see Fig. \ref{fig:spectrum}a). We consider bosonic atoms occupying the ground state of the respective potential well. Rydberg states are excited by a laser. The typical timescale associated with the fast (electronic) Rydberg dynamics is of the order of hundreds of nanoseconds. In this regime the external motion of the atoms is frozen and the time-evolution of a Rydberg gas is described by an arrangement of spin $1/2$ particles with Ising type interaction \cite{Weimer08,Raitzsch08,Sun08,Olmos09}. Here the two spin states correspond to the atomic ground and Rydberg state, and the laser detuning and its Rabi frequency give rise to an effective magnetic field.

Here we show that the above-mentioned setup allows to create and explore collective many-particle states that are entangled and extend over the entire lattice, e.g. spin waves, on a microsecond timescale. We consider a scenario in which the interaction with the laser is stronger than the interatomic interaction. This strong driving constrains the system's dynamics such that the collective excitations of the atomic ensemble are described by spinless fermions. We characterize these many-particle states and discuss how they can be excited from an initial state in which only ground state atoms are present. We finally show that the system offers the possibility to study fermions in the presence of a disorder potential although no external atomic motion takes place. Here the disorder is created by a spatially randomly varying Rabi frequency. We focus on a situation where this variation is caused by fluctuations of the atom number in each lattice site. These fluctuations are generated by quenching a superfluid gas of ground state atoms by a sudden increase of the lattice depth.

The Hamiltonian of a frozen Rydberg gas in a ring lattice with $L$ sites is given by
\begin{align}
  H_\mathrm{spin}=\sum_{k=1}^L\left[\Omega_k\sigma_x^{(k)}+\Delta\, \mathcal{P}_k+\beta\, \mathcal{P}_k \mathcal{P}_{k+1}\right]\label{eq:working_Hamiltonian}
\end{align}
with $\mathcal{P}_{k}=(1/2)(1+\sigma_z^{(k)})$ being the projector onto the
state in which a Rydberg atom is present on site $k$. $\sigma_x^{(k)}$ and
$\sigma_z^{(k)}$ are the Pauli spin matrices and
$\sigma_x^{(L+1)}=\sigma_x^{(1)}$. The first two terms of
Eq. (\ref{eq:working_Hamiltonian}) account for the coupling of the atoms to
the excitation laser within the rotating-wave approximation. If more than one
atom is present per lattice site a simultaneous excitation of two or more
Rydberg atoms is forbidden due to the large (interaction induced) energy shift
of this multiply excited state. This implies that in general, on each site
$k$, the excitation laser couples the product state $\left|P\right>_k
=\left[\left|g\right>_k\right]_1\otimes ... \otimes
\left[\left|g\right>_k\right]_{N_k}$ to the superatom state $\left|R\right>_k=
N_k^{-1/2}\mathcal{S}\left\{\left[\left|r\right>_k\right]_1\otimes
  \left[\left|g\right>_k\right]_2\otimes ... \otimes
  \left[\left|g\right>_k\right]_{N_k}\right\}$ ($\mathcal{S}$ is the
symmetrization operator, $N_k$ is the number of atoms contained in site $k$
and $\left|g\right>$ and $\left|r\right>$ are the single atom ground and
Rydberg states, respectively) \cite{Lukin01,Heidemann07}. Throughout, we
consider these two states to constitute a (super)atom located at site $k$,
and we define $\sigma_z^{(k)}\left|P\right>_k=-\left|P\right>_k$ and $\sigma_z^{(k)}\left|R\right>_k=\left|R\right>_k$.
They are coupled by the collective Rabi frequency  $\Omega_k=\sqrt{N_k}\Omega_0$, where $\Omega_0$ is the single-atom Rabi frequency. The detuning of the laser with respect to the transition $\left|P\right>_k\rightarrow \left|R\right>_k$ is $\Delta$. The last term of Hamiltonian (\ref{eq:working_Hamiltonian}) describes the interaction between (super)atoms located at different sites. We consider a regime in which the lattice spacing $a$ is much larger than the extension of the Wannier functions $\sigma$ (see Fig. \ref{fig:spectrum}a). The interaction strength between (super)atoms in neighboring sites is then given by $\beta=C_6/a^6$, with $C_6$ being the van-der-Waals coefficient. We consider only nearest neighbor interaction, which is well justified for sufficiently large rings since the next nearest neighbor interaction is a factor of $64\cos^6(\pi/L)$ smaller. Hamiltonians of the form (\ref{eq:working_Hamiltonian}) have been extensively and successfully used to study strongly interacting Rydberg gases \cite{Raitzsch08,Sun08,Olmos09,Weimer08,Schachenmayer09,Pohl09}.

Throughout this work we are interested in many-particle states that emerge in the limit of strong laser driving, i.e $\Omega_k\gg\beta$ and $|\Delta|\ll\Omega_k$. Since in this regime the first term of Eq. (\ref{eq:working_Hamiltonian}) dominates, it is convenient to diagonalize it applying
the unitary transformation $U=\prod_k \exp(-i (\pi/4) \sigma_y^{(k)})$. We then find $H=U^\dagger H_\mathrm{spin} U=H_\mathrm{xy}+H_1+H_2+\beta L/4$ with
\begin{eqnarray}
H_\mathrm{xy}&=& \sum_{k=1}^L\left[\Omega_k\sigma_z^{(k)}+\frac{\beta}{4}\!\left(\sigma_+^{(k)}\sigma_-^{(k+1)}\!+\sigma_-^{(k)}\sigma_+^{(k+1)}\right)\right]\label{eq:xy_model}\\
H_1&=&\frac{\Delta}{2}\sum_{k=1}^L \left(1-\sigma_x^{(k)}\right)\\
H_2&=&\frac{\beta}{4}\sum_{k=1}^L\left[\sigma_+^{(k)}\sigma_+^{(k+1)}+\sigma_-^{(k)}\sigma_-^{(k+1)}
-2\sigma_x^{(k)}\right],
\end{eqnarray}
where $H_\mathrm{xy}$ is the famous $xy$-model of a spin chain with a transverse magnetic field \cite{Lieb61}. After
the unitary transformation, the eigenstates of $\sigma^{(k)}_z$ are - in terms of the (super)atom states - given by $\left|\pm\right>_k=(1/\sqrt{2})U^\dagger\left[\left|P\right>_k\pm\left|R\right>_k\right]$ with $\sigma^{(k)}_z\left|\pm\right>_k=\pm\left|\pm\right>_k$.

Let us first consider the limit where each site of the lattice is occupied by the same number of atoms, i.e., $N_k=N_0$, which implies a constant Rabi frequency $\Omega_k=\Omega$. This is achieved, for example, if the lattice is initialized in a Mott-insulator state. For $\Omega\gg\beta$, the spectrum of $H$ decays into manifolds of nearly degenerate states having the same eigenvalue $m$ with respect to the operator $\sum_k \sigma^{(k)}_z$. These manifolds are separated by approximately $2\Omega$ (see Fig. \ref{fig:spectrum}b). As indicated in Fig. \ref{fig:spectrum}b, coupling between the manifolds is only caused by $H_1$ and $H_2$. The corresponding transition rates are approximately $\Delta^2/\Omega$ and $\beta^2/\Omega$ for $H_{1}$ and $H_{2}$, respectively. Hence, for sufficiently strong driving the system's dynamics is constrained to the $m$-manifolds and $H_{1,2}$ can be neglected.

The dynamics is then governed by the $xy$-model Hamiltonian (\ref{eq:xy_model}) which can be solved by introducing creation and annihilation operators of spinless fermions ($c^\dagger_k$ and $c_k$, respectively) through the Jordan-Wigner transformation $c_k=\exp\left(i\pi \sum_{j=1}^{k-1} \sigma^{(j)}_+\sigma^{(j)}_-\right)\,\sigma^{(k)}_-$ \cite{Lieb61}. One obtains $H_\mathrm{xy}=H_0+H_\mathrm{b}$ with
\begin{eqnarray}
  H_0=\sum_{k=1}^L\!\left[2\Omega_k\left(c^\dagger_kc_k-\frac{1}{2}\right)
  \!+\!\frac{\beta}{4}\left(c_k^\dagger c_{k+1}-c_k c_{k+1}^\dagger\right)\right]\label{eq:fermion_hamiltonian}
\end{eqnarray}
and $H_\mathrm{b}=-\frac{\beta}{4}\left(c_L^\dagger c_{1}-c_L c_{1}^\dagger\right)\left(e^{i\pi n_+}+1\right)$ being the boundary term due to the ring. The latter depends on the operator $n_+=\sum_{k=1}^L c^\dagger_kc_k$ and counts the number of fermions / (super)atoms in the state $\left|+\right>$. The creation operators $\eta^\dagger_{p,n}$ of the eigenexcitations of Hamiltonian (\ref{eq:fermion_hamiltonian}) can be found by a Fourier transform.
Due to the boundary term one has to distinguish between an odd ($p=o$) and even  ($p=e$)
number of fermions:
$\eta^\dagger_{p,n}=\frac{1}{\sqrt{L}}\sum_{k=1}^L \exp\left(i\alpha^p_n k\right)c^\dagger_k$ with $\alpha^\mathrm{o}_n=2n\pi/L$ and $\alpha^\mathrm{e}_n=2(n-1/2)\pi/L$.
The diagonal Hamiltonian reads
\begin{eqnarray}
  H^p_\mathrm{xy}=-L\Omega+\sum_{n=1}^L\eta^\dagger_{p,n}\eta_{p,n}\left(2\Omega+\frac{\beta}{2}\cos\alpha^p_n\right).\label{eq:fermion_hamiltonian_omega}
\end{eqnarray}
Its ground state reads $\left|G\right>=\prod_k \left|-\right>_k$ and the excited states are of the form $\left|ij\right>=\eta_{\mathrm{e},i}^\dagger\eta_{\mathrm{e},j}^\dagger\left|G\right>$ or $\left|ijk\right>=\eta_{\mathrm{o},i}^\dagger\eta_{\mathrm{o},j}^\dagger\eta_{\mathrm{o},k}^\dagger\left|G\right>$.
These states are eigenstates of the initial Hamiltonian (\ref{eq:working_Hamiltonian}) in the limit $\Omega_k=\Omega\gg\beta$.

\begin{figure}
\includegraphics[width=9.0cm]{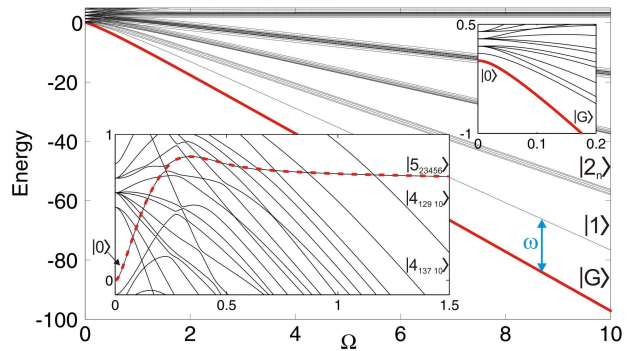}
\caption{Energy spectrum (fully symmetric states) of Hamiltonian (\ref{eq:working_Hamiltonian}) for a lattice with $10$ sites in units of $\beta$. Many-particle states can be accessed from the state $\left|0\right>$ by a temporal variation of the laser parameters. If $\Delta_0 >0$ (background figure and right inset [magnified view]) the state $\left|0\right>$ is adiabatically connected to the ground state $\left|G\right>$ of $H_\mathrm{xy}$ (thick orange line). See text for further explanation. If $\Delta_0 <0$ (left inset) the state $\left|0\right>$ cuts through a number of avoided crossings as $\Omega$ is increased. The dashed line shows a passage to the state $\left|5_{23456}\right>$. }\label{fig:fermion_spectrum}
\end{figure}

We now address the two questions: '\textit{Which states can be excited from a given initial state?}' and '\textit{How can they be excited?}'. To this end we define the initial conditions: i) The system shall be in a state in which no Rydberg atoms are present, i.e., $\left|0\right>=\prod_k \left|P\right>_k$. ii) The laser is turned off, i.e., $\Omega_0(0)=0$ and set to a certain detuning $\Delta(0)\equiv\Delta_0$. This corresponds to the left-hand side of the plots in Fig. \ref{fig:fermion_spectrum}, where we present the energy spectrum of Hamiltonian (\ref{eq:working_Hamiltonian}) ($L=10$) as a function of $\Omega$.

\textit{Which states can be excited?} - The symmetry properties of the system impose certain selection rules which imply that not all fermionic excitations are accessible from the initial state $\left|0\right>$ by a sweep of the laser parameters. In Ref. \cite{Olmos09} it was shown that Hamiltonian (\ref{eq:working_Hamiltonian}) and also (\ref{eq:fermion_hamiltonian_omega}) are invariant under cyclic shifts
($\mathcal{X}^\dagger\sigma^{(k)}_j \mathcal{X}=\sigma^{(k+1)}_j$) and reversal of the lattice sites ($\mathcal{R}^\dagger\sigma^{(k)}_j \mathcal{R}=\sigma^{(L-k+1)}_j$). Since the initial state is also invariant under the action of both operators, i.e., $\mathcal{X}^\dagger\left|0\right>=\mathcal{R}^\dagger\left|0\right>=\left|0\right>$, \emph{only states from this (fully symmetric) subspace are accessible} when $\Delta$ and $\Omega_0$ are varied.

Let us first study which single fermion states are actually accessible. A general state is given by $\eta^\dagger_{\mathrm{o},n}\left|G\right>$ where $n=1\dots L$. For a fully symmetric state we require $\eta^\dagger_{\mathrm{o},n}\!\left|G\right>=O^\dagger \eta^\dagger_{\mathrm{o},n}\!\left|G\right>=O^\dagger\eta^\dagger_{\mathrm{o},n}O O^\dagger \!\left|G\right>=O^\dagger\eta^\dagger_{\mathrm{o},n}O\!\left|G\right>$
where $O$ is a placeholder for $\mathcal{X}$ and $\mathcal{R}$. Here we have used the fact that $\left|G\right>$ - just like $\left|0\right>$ - is a fully symmetric state. Direct calculation shows that
$\mathcal{R}^\dagger \eta^\dagger_{\mathrm{o},n}\mathcal{R}=e^{i\frac{2\pi}{L}n}\eta^\dagger_{\mathrm{o},L-n}e^{i\pi n_+}$ and  $\mathcal{X}^\dagger \eta^\dagger_{\mathrm{o},n} \mathcal{X} = e^{-i\frac{2\pi}{L}n}\eta^\dagger_{\mathrm{o},n}e^{i\pi c^\dagger_1 c_1}+\!\frac{1}{\sqrt{L}}c_1^\dagger (e^{i\pi n_+}-1)$. The only fully symmetric state containing one fermion is thus $\left|1\right>=\eta^\dagger_{\mathrm{o},L}\left|G\right>$.  For states that contain two fermions one finds $\left|2_n\right>=\eta^\dagger_{\mathrm{e},L-n+1}\eta^\dagger_{\mathrm{e},n}\left|G\right>$
where $n=1\dots \lfloor L/2\rfloor$. In order to give a physical meaning to the states it is more instructive to use operators that refer to the atomic basis:
\begin{eqnarray*}
\left|1\right>&=&\frac{1}{\sqrt{L}}\sum_{k}\sigma_+^{(k)}\left|G\right>\\
\left|2_n\right>&=&\frac{2}{i\,L}\sum_{k>m}\sin\left[\frac{2\pi}{L}\left(n-1/2\right)\left(k-m\right)\right]\sigma_+^{(k)}\sigma_+^{(m)}\left|G\right>
\end{eqnarray*}
Hence, $\left|1\right>$ is a spin wave, or a superatom state that extends over the \textit{entire lattice} and the states $\left|2_n\right>$ are entangled two-atom states. Such excitations are of interest as they can yield a resource for the creation of single photons and photon pairs \cite{Porras08}. Excitations containing three fermions are constructed via
$\left|3_{lpq}\right>=\frac{1}{\sqrt{2}}\left[\eta^\dagger_{\mathrm{o},l}\eta^\dagger_{\mathrm{o},p}\eta^\dagger_{\mathrm{o},q}
-\eta^\dagger_{\mathrm{o},L-l}\eta^\dagger_{\mathrm{o},L-p}\eta^\dagger_{\mathrm{o},L-q}\right]\left|G\right>$, where the indices $l$, $p$ and $q$ can assume all distinct combinations obeying $l+p+q=\epsilon L$ with $\epsilon=1,2$. Further excited states are constructed in a similar fashion. A thorough investigation of their properties will be presented elsewhere.

\textit{How can the many-particle states be excited?} - We seek to selectively populate states in the regime $\Omega\gg\beta$. To this end we vary the laser parameters as a function of time, i.e., $\Delta\rightarrow \Delta(t)$ and $\Omega_0\rightarrow\Omega_0(t)$. Our starting point are the above-mentioned initial conditions (left-hand side of Fig. \ref{fig:fermion_spectrum}). We now consider two possible excitation methods which depend on the initial sign of the detuning:

In case of $\Delta_0<0$ (left inset of Fig. \ref{fig:fermion_spectrum}) the ground state at $\Omega=0$ does not coincide with the initial state $\left|0\right>$. With increasing $\Omega$ the energy of this state cuts through a number of avoided crossings. It is thus not a trivial task to find a proper trajectory $(\Delta(t),\Omega_0(t))$ which eventually connects $\left|0\right>$ with a desired eigenstate of $H_\mathrm{xy}$. Only in certain cases (see Refs. \cite{Pohl09,Schachenmayer09}) such a sequence can be guessed. A more general framework for finding a proper trajectory is provided by Optimal Control theory \cite{Peirce88}. Here certain constraints on the trajectory  can be imposed (e.g. to avoid too fast temporal oscillations) and the desired fidelity with which the final state is achieved can be set. This method is successfully applied to quantum information processing \cite{Calarco04}, molecular state preparation \cite{Somloi93} and optimization of number squeezing of an atomic gas confined to a double well potential \cite{Grond09}.

We pursue a different route and set $\Delta_0>0$ at $\Omega=0$. As shown in the right inset of Fig. \ref{fig:fermion_spectrum}, $\left|0\right>$ now coincides with the ground state.
Ramping up $\Omega$ this state is adiabatically connected to the ground state
$\left|G\right>$ of $H_\mathrm{xy}$. For small $\Omega$ the distance to
adjacent states can be increased by increasing $\Delta$ such that
non-adiabatic transitions are strongly suppressed. From the state
$\left|G\right>$ many-particle states can be excited by introducing an
oscillating detuning $\Delta(t)=\Delta_\mathrm{osc}\cos(\omega t)$ giving rise
to a time-dependent $H_1$. Within the rotating-wave approximation the coupling
matrix element between $\left|G\right>$ and the fully symmetric single fermion
state is
$\left<G\right|H_1\left|1\right>=-(\Delta_\mathrm{osc}\sqrt{L})/4$. The
oscillating detuning acts like a 'laser' which drives transitions between
different many-particle states. Hence, if the (radiofrequency) oscillation of
the detuning is on resonance, i.e., $\omega \approx \omega_L=2\Omega+\beta/2$
(see Fig. \ref{fig:fermion_spectrum}), Rabi flops between the ground state and
the first excited state are performed. Resonance with the
$\left|1\right>\rightarrow \left|2_n\right>$ transitions can be avoided by
choosing a not too high value of $\Omega$. Here, level shifts due to $H_2$,
which are of the order $\beta^2/\Omega$, bring such unwanted transitions out
of resonance. This offers also a possibility to excite higher lying states by
first performing a $\pi$-pulse on the $\left|G\right>\rightarrow
\left|1\right>$ transition
followed by a $\pi$-pulse with frequency adapted to
the transition $\left|1\right>\rightarrow\left|2_n\right>$.

Due to the limited lifetime of the Rydberg atoms (e.g. 66 $\mu s$ for Rb in the 60s state), the whole excitation procedure has to take place on a $\mu s$ time scale. Once prepared, the many-particle states are mapped to a stable state by performing a global $\pi$-pulse on the single-atom transition $\left|r\right>\rightarrow \left|s\right>$, where $\left|s\right>$ is a stable storage state. Experimentally, the excited many-particle states can be detected by counting the number of atoms in the $\left|+\right>$-state which  corresponds to the number of fermions.

\begin{figure}
\includegraphics[width=9.3cm]{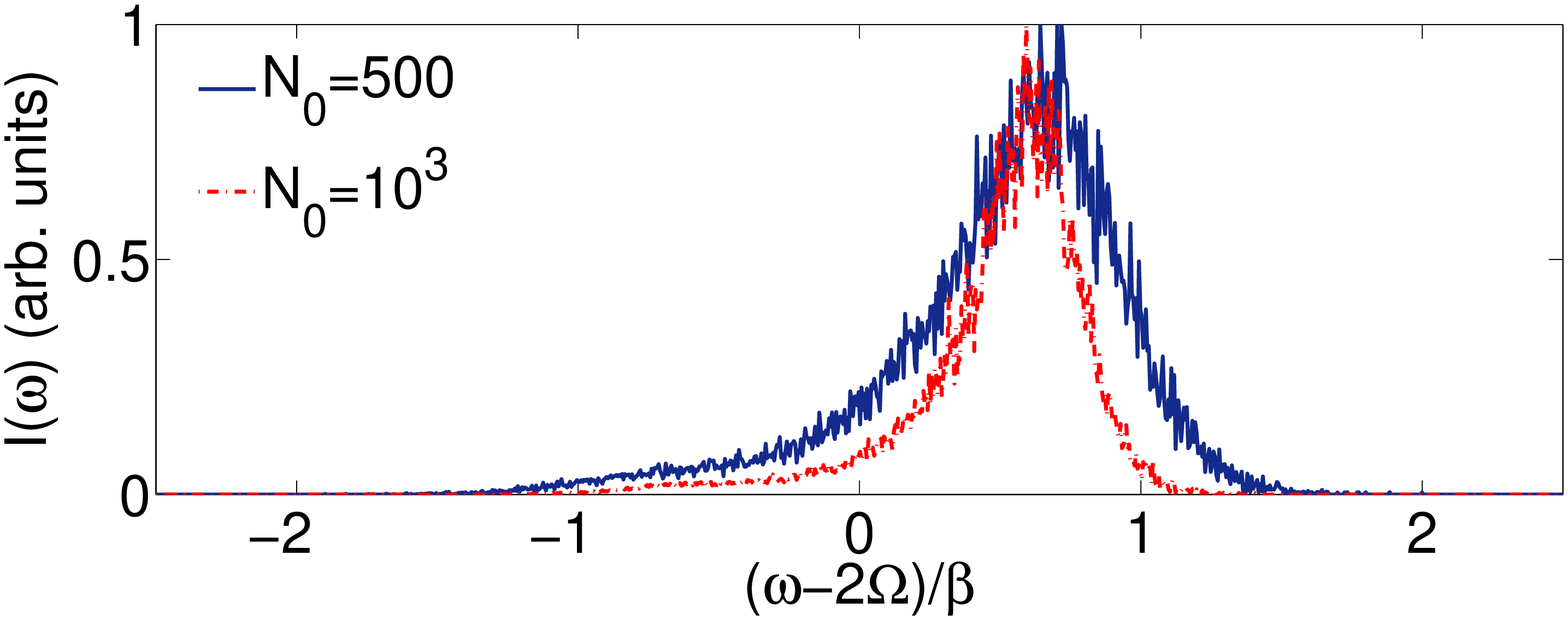}
\caption{Absorption profile (excitation probability of a single fermion state) for the $\left|G\right>\rightarrow \left|1\right>$ transition for two strengths of disorder, $\Omega/\beta=10$ and $L=50$. The disorder strength is controlled by the mean number of atoms $N_0$ per site. The results are averages over $1000$ realizations.}\label{fig:profile}
\end{figure}

So far we have assumed a constant $\Omega_k$. We will now consider a situation
in which $\Omega_k$ is not constant but fluctuates randomly around a mean
value, i.e. $\Omega_k=\Omega+\delta\Omega_k$.
In Eq. (\ref{eq:fermion_hamiltonian}) the fluctuating part
$\delta\Omega_k$ introduces a random single particle potential for the fermions, and  gives rise to the Hamiltonian of Anderson localization \cite{Anderson58}. Hence, a lattice gas of Rydberg atoms offers the possibility to study fermions in a disorder potential although no external atomic motion takes place. The required spatial variation of $\Omega_k$ can, for instance, be achieved by a speckle potential or standing waves with incommensurate frequencies \cite{Billy08,Roati08}. We present an alternative route: We start in a situation in which $N_\mathrm{g}$ ground state atoms are prepared in a superfluid state $\left|\mathrm{SF}\right>=(N_\mathrm{g}!\,L^{N_g/2})^{-1}\left[\sum_{k=1}^L b_k^\dagger\right]^{N_\mathrm{g}}\left|\mathrm{vac}\right>$ where $b^\dagger_k$ creates a ground state atom in state $\left|g\right>$ at site $k$. Disorder is introduced by a quench of this superfluid through a sudden increase of the depth of the lattice potential. For a sufficiently large number of atoms $N_\mathrm{g}$ the Rabi frequency of site $k$ is then given by $\Omega_k=\Omega_0\sqrt{N_k}=\sqrt{N_0}\Omega_0+\delta\Omega_k=\Omega+\delta\Omega_k$
with $\delta\Omega_k=\delta N_k\, \Omega/(2 N_0)$ and $N_0=N_\mathrm{g}/L$. Here $\delta\Omega_k$ is a random variable with $\left<\delta\Omega_k \right>=0$ and
$\left<\delta\Omega_k\, \delta\Omega_m\right>=\Omega^2/(4N_0)\left[\delta_{km}-L^{-1}\right]$.
The disorder destroys the symmetry properties of the system and hence also the
selection rules for transitions between many-particle states. As a
consequence, there are now in general $L$ single fermion states (instead of
$1$) accessible from the state $\left|G\right>$ when an oscillating detuning
is applied. Instead of a single sharp line at $\omega_L$ the (averaged)
absorption profile of the $\left|G\right>\rightarrow
\left|1\right>$-transition broadens and becomes asymmetric,
see Fig. \ref{fig:profile}. For small disorder, one finds from second order perturbation theory, that the red wing is approximated by $I(\omega)\sim \Omega^2/(4N_0|\omega-\omega_L|^2)$.
This disorder-induced line broadening can be detected by counting the number of $\left|+\right>$-atoms as a function of the excitation frequency $\omega$.

In our considerations we have assumed that the atoms are strongly localized, i.e., $a\gg\sigma$. In practice there is a finite width of the wave-packet, caused by the uncertainty principle and finite temperature. This will lead to disorder also in the interaction energy $\beta$, which can be also treated in the present framework but is beyond the scope of this work. Moreover, the number of atoms in a single site cannot be made too large due to three-body losses. Bearing this in mind, the system opens exciting perspectives for creating complex many-particle states with interesting prospects for the study of disorder and the generation of non-classical light.

\begin{acknowledgments}
B.O. and  R.G.F. acknowledge the grants FIS2008--02380 (MICINN),  FQM-207 and  FQM-2445 (JA), and  B.O. the support of MEC under the program FPU.
\end{acknowledgments}

\end{document}